\documentclass[%
 reprint,
 amsmath,amssymb,
 aps,
floatfix,
]{revtex4-1}
\usepackage{float}
\usepackage{graphicx}% Include figure files
\usepackage{dcolumn}% Align table columns on decimal point
\usepackage{bm}
\usepackage{mathtools}
\usepackage{caption}
\usepackage{subcaption}
\usepackage{bbm}
\usepackage{verbatim}
\usepackage[colorlinks = true, linkcolor = blue, urlcolor = blue, citecolor = red, anchorcolor = blue]{hyperref}
\usepackage[normalem]{ulem}%For editing. \sout for strikeout

\usepackage{xcolor}

\begin{document}

\preprint{APS/123-QED}

\title{Measuring two temperatures using a single thermometer}% Force line breaks with \\

\author{Harshit Verma}
 \email{h.verma@uqconnect.edu.au}
 \affiliation{Eviden, Blk 988 Toa Payoh North, \#08-01, Singapore 319002}

\author{Fabio Costa}
\affiliation{Nordita, Stockholm University and KTH Royal Institute of Technology, Hannes Alfv{\'e}ns v{\"a}g 12, SE-106 91 Stockholm, Sweden}
 \affiliation{Australian Research Council Centre of Excellence for Engineered Quantum Systems (EQUS),
School of Mathematics and Physics, The University of Queensland, St Lucia, QLD 4072, Australia}

 \email{h.verma@uq.edu.au}

\date{\today}% It is always \today, today,
             %  but any date may be explicitly specifiedq   
\begin{abstract}
We consider the question: Is it possible to measure two temperatures simultaneously using a single thermometer? Under common circumstances, where the thermometer can interact with only one bath at a time and the interaction leads to complete thermalization, this is clearly impossible because the final state of the thermometer would be independent of the temperature of the first bath.
In this work, we show that this task can indeed be accomplished with the assistance of quantum control. In particular, we consider a composite particle with multiple quantum degrees of freedom (DoF) as a temperature sensor, where one of the DoF -- termed as internal DoF -- is susceptible to the local temperature, thereby functioning as a thermometer, whereas another DoF -- termed external DoF -- is quantum-controlled. We leverage the entanglement between the aforementioned DoF in a composite particle for two-temperature thermometry by preparing the external DoF in a quantum superposition, exposing the internal DoF to two local temperatures. We show that such a particle used in a Mach-Zehnder type interferometer, or a quantum switch -- which allows quantum control over the order of application of quantum channels -- can be used to estimate two temperatures simultaneously, thus affirming our main proposition. For each of these setups, we obtain the variance in the estimated temperatures through the multi-parameter Cram\'er-Rao bound, and compare their performances based on the range of total variance of the two temperatures estimated. On benchmarking all the setups based on the total variance of the estimated temperatures, we find that a quantum switch with a qudit probe outperforms other setups. On restricting our probe to be a qubit, we find that quantum switch performs equally well as a Mach-Zehnder type interferometer. 

\end{abstract}

\maketitle

\section{Introduction}
Since the inception of quantum technology, quantum metrology has been one of its prominent streams spanning both domains -- theoretical and experimental. Traditionally, quantum metrology has been associated with developing techniques and schemes for better sensing and estimating quantities of physical relevance. Many such quantities, such as magnetic field and temperature, are of practical and fundamental interest, but cannot be measured directly in the quantum regime because they do not correspond to quantum observables. Nevertheless, it has been widely shown that such quantities can be estimated using other quantum observables, i.e., indirectly. To underline, the central question of quantum metrology is: How to estimate quantities which cannot be directly observed and also, how well can such quantities be estimated \cite{Helstrom1967,Helstrom1969,PhysRevLett.72.3439,Holevo2001}?

A common layout of a quantum sensing experiment consists of the following parts, generally carried out in the order below:
\begin{itemize}
    \item Probe preparation: The quantum probe is initialized in some state.
    \item Probe interaction: The probe interacts with the environment to encode some information about the relevant parameter in its state.
    \item Probe Measurement: Measurements are performed on the probe to access the information/parameter encoded.
    \item Estimation: Using the information obtained in the measurement stage to construct a probability distribution for the parameter of interest using an estimator.
\end{itemize}
An estimator functional is defined as a mapping from the space of measurement outcomes to the space of parameter of interest. The answer to the question of how well can a parameter be estimated lies in identifying the optimal estimator, which leads to a qualitative best estimate of the parameter. In the classical estimation theory, and for a single real parameter $\lambda$, an optimal estimator must saturate the classical Cram\'er-Rao bound (CCRB) \cite{Paris2009} given as follows:
\begin{align}
    V(\lambda) = \frac{1}{N {\mathbb{F}}(\lambda)}~,
    \label{eq:ccrb}
\end{align}
where $N$ is the number of repetitions of the estimation procedure, $\mathbb{F}(\lambda)$ is the classical Fisher information, and $V(\lambda)$ is the mean-square error in the estimator functional. For unbiased estimators, the latter is equal to the variance of the parameter lambda, i.e. $V (\lambda) = \operatorname{Var}(\lambda)$. Note that the scaling factor $N$ arises from the central limit theorem, and is essential in the classical as well as quantum versions of the Cram\'er-Rao bound. It is handy to express Eq.~\eqref{eq:ccrb} in terms of the precision in the estimated quantity ($\delta \lambda = \sqrt{V(\lambda)} $) as follows:
\begin{align}
    \delta \lambda = \frac{1}{\sqrt{N} \sqrt{{\mathbb{F}}(\lambda)}}~,\nonumber
\end{align}
where the LHS provides a quantitative metric of the estimation procedure i.e. a good procedure should minimize the precision.

\begin{figure*}
    \centering
    \includegraphics[scale = 1.2]{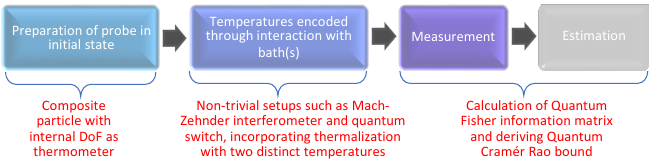} 
    \caption{Sequence of operations in the estimation scheme. We focus on the probe interaction step where we propose to use various setups based on Mach-Zehnder interferometer and quantum switch, alongside thermalization maps corresponding to two distinct temperatures.}
    \label{fig:enc}
\end{figure*}

A similar bound is also valid for quantum estimation problems wherein the classical Fisher information in Eq.~\eqref{eq:ccrb} is replaced by the quantum Fisher information to obtain the quantum Cram\'er-Rao bound (QCRB) \cite{Paris2009}. Since the scaling with the repetitions of the experiment remains the same -- on account of its basis on the central limit theorem -- this is termed as the standard quantum limit (SQL). The field of quantum enhanced metrology demonstrates a theoretical revision in the SQL, with an improvement in the precision bound by a factor of $\frac{1}{\sqrt{N}}$, which is termed as the Heisenberg limit \cite{Giovannetti2006}. Note that the achievability of QCRB is conditional on the usage of purely quantum resources such as entanglement \cite{Ramanathan2005} and squeezing. Many recent works in quantum metorology are focused on achieving the Heisenberg limit, particularly in noisy scenarios which inhibits the role of the aforementioned quantum resources \cite{Escher2011,Demkowicz,Zhou2018}. From the laboratory setup outlined before, this could translate to optimization in preparation \cite{PhysRevLett.113.250801}, interaction \cite{PhysRevA.65.025802, Hou2019, PhysRevLett.109.233601} and measurement phases \cite{Meyer2021, Ho2021}. 

So far, we have considered the salient features of an estimation problem encapsulating a single parameter. The generalization to multiple parameters \cite{Albarelli2020}, which is necessary for our study, is discussed later in Sec.~\ref{sec:th} and involves subtle, but important differences \cite{Szczykulska2016,Liu_2019}.

Temperature is an extensive property of profound fundamental significance in both classical and quantum thermodynamics \cite{thermo_rev}. We will be mostly concerned with the estimation of temperature or thermometry \cite{PhysRevLett.114.220405,PhysRevA.82.011611,DePasquale2016,DePasquale2018, PhysRevA.107.042614, Abiuso_2024} and naturally, quantum metrology forms the backbone for this task \cite{PhysRevA.83.012315,PhysRevResearch.2.033389}. The task that we consider here falls in the paradigm of local quantum thermometry wherein it is assumed that the broad range of the parameter of interest is known, in contrast with global thermometry wherein no prior information about the parameter is assumed. Several works have proposed new ways to perform low-temperature thermometry \cite{Wu1998, Potts2019,PhysRevA.91.012331,PhysRevA.96.012316,Hofer2017,Campbell2017, Mukherjee2019,Seah2019,Mathias2020,Glatthard2022}, in line with the generic efforts in quantum metrology to increase the precision, or achieve the theoretical limit in experiments, as outlined before.

%Moreover, we shall be confined to low temperature thermometry, which has been found to be particularly difficult \cite{Wu1998,Paris2016,Potts2019}. Therefore, though the low temperature regime dominates our discussion, nonetheless, the methods introduced in this manuscript are also relevant for other temperature regimes.

For a single thermometer to be susceptible to two temperatures, one may deem a setup with successive interactions of the thermometer with baths at distinct temperature to be useful. In a scenario where the interaction of the probe with the bath leads to complete thermalization of the probe, this setup is inferred to be trivially unsuitable for the task because the probe only carries information about the temperature of the bath involved in the last interaction. 

In this manuscript, we tackle the question of estimating two temperatures simultaneously using a single thermometer and answer it in an affirmative. To this end, we model the thermometer as a quantum system, the thermalization process as a quantum channel that brings the thermometer to a thermal state, and we include an additional quantum system that controls how the thermometer interacts with the two baths. This approach aligns closely to models of clocks travelling in superpositions of trajectories \cite{Zych2011, Verma2021}, effectively extending the analysis of quantum control of proper-time measurements to quantum control of temperature measurements. We focus on using the above setups in the probe interaction step, and analyse if it could enable multi-parameter sensing, particularly two-temperature sensing (see Fig.~\eqref{fig:enc}). The exposure of the quantum probe or thermometer to a particular temperature is modelled as an interaction with a bath at a finite temperature, and in the open quantum systems framework, corresponds to the application of a thermalizing map on the probe.

At first, we consider a composite particle in a Mach-Zehnder interferometer. In a composite particle -- which possesses multiple DoF -- the internal DoF is exposed to bath(s) and acts as a thermometer, whereas the external DoF is quantum controlled -- allowing interaction with two baths at distinct temperatures or a single bath whose effective temperature is also quantum-controlled. These schemes effectively implement the definitions of temperature superposition proposed in \cite{Wood2021}. We develop this model and delve into the intricacies of such a setup in Sec. \ref{sec:sense}.

We also consider the quantum switch, a process that allows quantum control over the order of application of two quantum channels \cite{PhysRevLett.121.090503}, here two thermalizing channels corresponding to distinct temperatures. Intuitively, such a setup seems to be an appropriate candidate for the two-temperature thermometry task that we have identified.

Using the above setups, we obtain the bounds on the precision of estimated temperatures, using the multi-parameter Cram\'er-Rao bound. Therefore, we assert that it is indeed possible to measure two temperatures using a single thermometer, albeit with the assistance of quantum control.

In Sec.~\ref{sec:th}, we discuss the general framework of multi-parameter estimation, the corresponding precision bounds, thermalization maps, and the essentials of quantum processes: a framework encompassing the quantum switch. Thereafter, in Sec.~\ref{sec:sense} we consider the estimation task with two thermalizing maps in different evolutionary setups mentioned before, one by one and outline the key results, including the numerically calculated precision bounds, and their attainability. Finally, in Sec.~\ref{sec:conc}, we write the concluding remarks, while also comparing the results obtained for different setups. In this section we also highlight the future outlook and research that this work could engender.

\section{Theoretical Framework}\label{sec:th}

\subsection{Cram\'er-Rao bound for multi-parameter estimation}
The primary problem that we address here is of estimating the variable (vector) representing the temperatures: $\vec{T}=[T_1,T_2]^\textrm{\textbf{T}}$ encoded in the parametric density operator generically denoted by $\rho_T$. To recapitulate, the thermometry procedure includes the preparation of an input state of the probe, parameterization (encoding the parameter in the state) through application of thermalization channel(s), quantum measurement (POVM), and classical estimation of the parameter from the measurement data. Hence, the $\rho_T$ mentioned above is the state of the probe after the parameterization step, which is sufficient to calculate the theoretical bounds, as we see later. In the data processing stage, a classical estimator function is used to envisage a cost function, which is aimed to be minimized. Similar to the role of quantum Fisher information in the single parameter case, in a multiparameter estimation problem, the quantum Fisher information matrix (QFIM) is used to provide a bound on the uncertainties, specifically the mean square error matrix (MSEM) of the estimator. For a locally unbiased estimator, the bound -- called the (Quantum) Cram\'er-Rao bound is given in the terms of the covariance matrix of the variable ($\vec{T}$), as follows \cite{Liu_2019, mihailescu2023multiparameter}:
\begin{eqnarray}
Cov(\vec{T}) \geq \frac{\mathcal{Q}_{\boldsymbol{T}}^{-1}}{N}~,
\label{eq:CRB}
\end{eqnarray}
where $N$ is the number of times the experiment is repeated, and $\mathcal{Q}_{\boldsymbol{T}}$ is the QFIM, which is only valid for non-singular QFIM. The QFIM mentioned here is derived using the symmetric logarithmic derivative (SLD) of the parametric density matrix $\rho_T$, which is given as follows:
\begin{eqnarray}\label{SLD}
\partial_r\rho_T = \frac{L_r\rho+\rho L_r}{2}~,\nonumber
\end{eqnarray}
where $r \in \vec{T}$ is the parameter under consideration and $L_r$ the corresponding SLD operator. As such, the entries of QFIM are given as follows:
\begin{eqnarray}
\mathcal{Q}^{nm} = \frac{1}{2}\operatorname{Tr}(\rho_T \{L_n,L_m\})
\end{eqnarray}
where $\{.,.\}$ denotes the anti-commutator. It is known that for the parametric density matrix $\rho_T = \sum_{\sigma_i \in \mathcal{S}} \sigma_i |\sigma_i\rangle\langle\sigma_i| $, where $\mathcal{S}$ is the support, the entries of QFIM can be written as follows \cite{Liu_2019}:
\begin{eqnarray}
\mathcal{Q}_{\boldsymbol{T}}^{n m} &=\sum\limits_{\sigma_{i} \in \mathcal{S}} \frac{\left(\partial_{n} \sigma_{i}\right)\left(\partial_{m} \sigma_{i}\right)}{\sigma_{i}}+\sum\limits_{\sigma_{i} \in \mathcal{S}} 4 \sigma_{i} \operatorname{Re}\left(\left\langle\partial_{n} \sigma_{i} \mid \partial_{m} \sigma_{i}\right\rangle\right) \nonumber \\
&-\sum\limits_{\sigma_{i}, \sigma_{j} \in \mathcal{S}} \frac{8 \sigma_{i} \sigma_{j}}{\sigma_{i}+\sigma_{j}} \operatorname{Re}\left(\left\langle\partial_{n} \sigma_{i} \mid \sigma_{j}\right\rangle\left\langle\sigma_{j} \mid \partial_{m} \sigma_{i}\right\rangle\right)~,
\label{eq:QNM}
\end{eqnarray}
where $\operatorname{Re}$ denotes the real part. It is worth mentioning here that the bound in Eq.~\eqref{eq:CRB} may not be necessarily attainable i.e. there might not be a set of POVMs that saturate the bound. For attainability i.e. saturation of the bound, the following condition must be satisfied \cite{Liu_2019}:
\begin{eqnarray}
\operatorname{Tr}\left(\rho_T [ L_n,L_m ] \right) = 0~ \forall~n,m~\in~\vec{T}~~.
\label{eq:attain}
\end{eqnarray}
Explicitly, the SLD operators are written in the basis of eigenvectors of $\rho_T$ as follows:
\begin{align}
    \langle \sigma_{i}|L_{m}| \sigma_{j}\rangle=\delta_{i j} \frac{\partial_{m} \sigma_{i}}{\sigma_{i}}+\frac{2(\sigma_{j}-\sigma_{i})}{\sigma_{i}+\sigma_{j}}\langle \sigma_{i} | \partial_{m} \sigma_{j}\rangle~,
\end{align}
where $|\sigma_{i}\rangle, |\sigma_j\rangle$ are eigenvectors corresponding to the eigenvalues $\sigma_i, \sigma_j  \in \mathcal{S}$ respectively. For $|\sigma_i\rangle \in \mathcal{S}, |\sigma_j\rangle \notin \mathcal{S}$, $\langle \sigma_{i}|L_{m}| \sigma_{j}\rangle=-2\langle \sigma_i|\partial_{m} \sigma_j \rangle$, and for $\sigma_{i,j}\notin\mathcal{S}$, $\langle \sigma_{i}|L_{m}| \sigma_{j}\rangle=0$ \cite{Liu_2019}.
Therefore, in all the cases that we consider, we will check whether the above condition is satisfied in the estimation task.
 Note that Eq.~\eqref{eq:CRB} should be interpreted as:
\begin{eqnarray}
Cov(\vec{T})- \frac{\mathcal{Q}_{\boldsymbol{T}}^{-1}}{N} \geq 0~, 
\label{eq:QCRB_der}
\end{eqnarray}
namely as the condition that the matrix on the left hand side is positive semi-definite. In the two parameter estimation task that we have considered, $\mathcal{Q}$ is of the following type:
\begin{align}
\mathcal{Q}_{\boldsymbol{T}}=\left(\begin{array}{cc}
\mathcal{Q}_{T_1 T_1} & \mathcal{Q}_{T_1 T_2} \\
\mathcal{Q}_{T_1 T_2} & \mathcal{Q}_{T_2 T_2} \nonumber
\end{array}\right)
\end{align}
for which, the inverse, if it exists, is of the following form: 
\begin{align}
\mathcal{Q}^{-1}_{\boldsymbol{T}}=\frac{1}{\operatorname{det}(\mathcal{Q_{\boldsymbol{T}}})}\left(\begin{array}{cc}
\mathcal{Q}_{T_2 T_2} & -\mathcal{Q}_{T_1 T_2} \\
-\mathcal{Q}_{T_1 T_2} & \mathcal{Q}_{T_1 T_1} \nonumber
\end{array}\right). 
\end{align}
As for the covariance matrix, it can be expressed as: $Cov(\vec{T})=\left(\begin{array}{cc}
\operatorname{Var}(T_1) & \operatorname{Cov}(T_1,T_2) \\
\operatorname{Cov}(T_1,T_2) & \operatorname{Var}(T_2)
\end{array}\right)$, where $\operatorname{Cov}$ refers to the covariance and $\operatorname{Var}$ refers to the variance. 

It is useful to note that, for any real and symmetric matrix $Z =\left(\begin{array}{cc}
a & b \\
b & c
\end{array}\right)$, $Z \geq 0$ is equivalent to stating that both the eigenvalues of $Z$ are positive, which is true if and only if $\operatorname{Tr} (Z) \geq 0$ and $\operatorname{Det} (Z) \geq 0$. In terms of the matrix elements, these conditions read $a + c\geq 0$ , $ac \geq b^2$. Since $b^2 \geq 0$ by itself, it follows that $ ac \geq 0$, which, taken with the condition $a + c\geq 0$, can be true only if $a \geq 0, ~\text{and}~ c\geq 0$ hold individually.
We use these refined inequalities on the matrix elements, arising from the semi-positivity of the matrix $\big(Cov(\vec{T})- \frac{\mathcal{Q}_{\boldsymbol{T}}^{-1}}{N}\big)$, noting that it is also a real and symmetric matrix.

Therefore, the matrix inequality given in Eq.~\eqref{eq:QCRB_der} can be reduced to the following set of inequalities:
\begin{eqnarray}
\operatorname{Var}(T_1)~\geq &~&\frac{\mathcal{Q}_{T_2 T_2}}{N~\operatorname{det}(\mathcal{Q_{\boldsymbol{T}}})}\nonumber\\
\operatorname{Var}(T_2)~\geq &~&\frac{\mathcal{Q}_{T_1 T_1}}{N~\operatorname{det}(\mathcal{Q_{\boldsymbol{T}}})}\nonumber\\
\bigg(\operatorname{Var}(T_1) - \frac{\mathcal{Q}_{T_2 T_2}}{N~\operatorname{det}(\mathcal{Q_{\boldsymbol{T}}})}\bigg)&&\bigg(\operatorname{Var}(T_2) - \frac{\mathcal{Q}_{T_1 T_1}}{N~\operatorname{det}(\mathcal{Q_{\boldsymbol{T}}})}\bigg) \geq \nonumber \\
\bigg[\operatorname{Cov}(T_1,T_2) &+& \frac{\mathcal{Q}_{T_1 T_2}}{N~\operatorname{det}(\mathcal{Q_{\boldsymbol{T}}})} \bigg]^2
\label{eq:2parameterbounds}
\end{eqnarray}
If the attainability condition in Eq.~\eqref{eq:attain} is satisfied, the following expressions are obtained:
\begin{eqnarray}
\operatorname{Var}(T_1)~= ~\frac{\mathcal{Q}_{T_2 T_2}}{N~\operatorname{det}(\mathcal{Q_{\boldsymbol{T}}})}\nonumber\\
\operatorname{Var}(T_2)~= ~\frac{\mathcal{Q}_{T_1 T_1}}{N~\operatorname{det}(\mathcal{Q_{\boldsymbol{T}}})}\nonumber\\
\operatorname{Cov}(T_1,T_2) = - \frac{\mathcal{Q}_{T_1 T_2}}{N~\operatorname{det}(\mathcal{Q_{\boldsymbol{T}}})} 
\label{eq:bnds}
\end{eqnarray}
\begin{figure}
    \centering
    \includegraphics[scale=0.8]{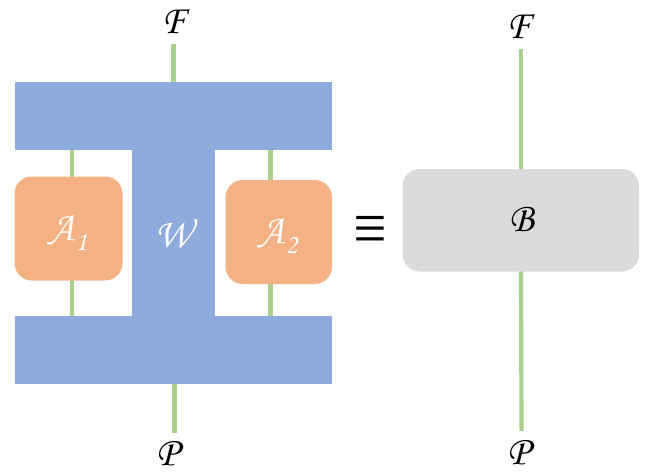}
    \caption{Illustrative example of a bipartite quantum supermap acting on two maps ($\mathcal{A}_i s$), represented by the resultant map ($\mathcal{B}$) on the right. $\mathcal{P}$, and $\mathcal{F}$ denote the input and output Hilbert spaces of the resultant quantum channel $\mathcal{B}$. The input state defined on $\mathcal{P}$ is affected by the process and the maps (equivalent to application of the channel $\mathcal{B}$) -- which encodes the requisite parameters in the state, and the output state defined on $\mathcal{F}$ is used in the estimation task for calculating the bounds.}
    \label{fig:process}
\end{figure}
We will use the above expressions to calculate the minimum variance that can be achieved in setups based on a composite particle in Mach-Zehnder interferometer and quantum switch. The attainability condition is tested for each setup to ensure that the inequalities in Eq.~\eqref{eq:2parameterbounds} are saturated. 

For a multi-parameter task such as ours, the total variance, i.e., $\operatorname{Var}(T_1) + \operatorname{Var}(T_2)$, is another figure of merit, which lets us compare various setups. For the two-temperature estimation task, it is bound by the trace of the inverse of the QFIM, written mathematically as follows \cite{Liu_2019}:

\begin{eqnarray}
   \operatorname{Var}(T_1) + \operatorname{Var}(T_2) \geq \frac{1}{N} \operatorname{Tr} (\mathcal{Q}_{\boldsymbol{T}}^{-1})
   \label{eq:tot_var}
\end{eqnarray}

We will evaluate the total variance for various setups in Section \ref{sec:sense}, and compare their performance using this metric.

%It is important to mention here that there exists a class of inequalities like the one in Eq.~\eqref{eq:CRB}. E.g. the Holevo Cram\'er Rao bound \cite{PhysRevLett.123.200503} is defined on the right logarithmic derivative (RLD) instead of SLD. Though such inequalities are of general interest, they may be equivalent to one another when the attainability criteria given in Eq.~\eqref{eq:attain} is satisfied, which we show to be true in all the cases that we have considered.

\subsection{Thermalization}

Measuring the temperature in any setup must include an interaction of the probe with a bath at a finite temperature. Such a bath can be considered to be in a Gibbs state given as: $\rho_G = \frac{e^{-\beta H_B}}{\mathcal{Z}}$, where $\beta = \frac{1}{K_B T}$ is the inverse temperature, $H_B = \sum_i E_i |i \rangle \langle i|$ is the bath Hamiltonian, and $\mathcal{Z}$ is the partition function. Essentially, the probe must thermalize with the bath to assume the same temperature. In this manuscript, we set the Boltzamnn constant to be unity i.e. $K_B = 1$ , and the temperature is assumed in the unit corresponding to this assumption.

Such an interaction step can also be denoted as the application of a quantum channel -- thermalization map -- on the probe. These maps are known as thermal attenuators for finite dimensional systems, and are used in modelling thermal effects on such quantum systems \cite{Rosati2018}. In most of our setups, we consider a qubit as a probe with the generalized amplitude damping channel (GADC) which acts as a thermalization map \cite{PhysRevA.102.012401,Chessa2021}. 

In the channel formalism, GADC is a qubit to qubit channel ($\mathcal{E}$) which is represented  as: $\mathcal{E}(\rho) = \tilde{\rho}$ where the input ($\rho$) and output ($\tilde{\rho}$) have Hilbert spaces of the same dimension.
The action of GADC on a probe can also be expressed in the Kraus formalism as $\mathcal{E}(\rho) = \sum_k R_k \rho R_k^{\dagger}$, where $R_k$ are the Kraus operators corresponding to the channel $\mathcal{E}$. The Kraus operators for the qubit GADC are listed below:
\begin{eqnarray}
\mathrm{R}_{1}&=&\sqrt{p}~( |0\rangle\langle 0| + \sqrt{1-\eta} |1\rangle\langle 1|), \nonumber\\
\mathrm{R}_{2}&=&\sqrt{1-p}~(\sqrt{1-\eta} |0\rangle\langle 0| + |1\rangle\langle 1|), \nonumber\\
\mathrm{R}_{3}&=&\sqrt{p \eta}~|0\rangle\langle 1|, \nonumber\\
\mathrm{R}_{4}&=&\sqrt{(1-p)~\eta } |1\rangle\langle 0|,
\label{eq:kraus}
\end{eqnarray}
with $0 \leq \eta, p \leq 1$. The above Kraus operators incorporate the temperature of thermalization (bath) as the inverse temperature $ \beta = \log_{2}\left(\frac{p}{1-p}\right)$.  $\eta$ is interpreted as the strength of the thermalizing map, and if we set $\eta=1$, a single application of the map is suitable to thermalize the probe completely.

The application of such a thermalizing map/quantum channel can be equivalently modelled using a probe-bath unitary interaction operator. In the pure state formalism, we can assume the bath to be in a purified state of the Gibbs state. Such a state is constituted by (at least) two qubits, and can be written in the energy eigenbasis as follows:

\begin{eqnarray}
\left|\theta^{\beta}\right\rangle \equiv \frac{e^{-E_0\beta/2}}{\sqrt{e^{-E_0\beta}+e^{-E_1\beta}}}|0,0\rangle+\frac{e^{-E_1\beta/2}}{\sqrt{e^{-E_0\beta}+e^{-E_1\beta}}}|1,1\rangle~\nonumber\\
\label{eq:bath}
\end{eqnarray}
where the bath Hamiltonian is $H_B = H_P \otimes \mathbb{I}_2 + \mathbb{I}_2 \otimes H_P$, where $H_P = \sum_i E_i|i\rangle\langle i|$. While the introduction of purified bath would seem unnecessary at this point, the pure state formalism is required in one of the setups which construes ``operational meaning of superposition of temperatures" introduced in \cite{Wood2021} and it is one of the setups that we have used in the two-temperature estimation task. It is also useful to note that such a dilation consisting of two qubits is not unique; the details about the general case can be found in \cite{Wood2021}.

The unitary operator for the thermalization of the probe, governing the interaction between the probe and one of the qubits of the purified bath is given as follows:
\begin{eqnarray}
U^{\eta}_{PB}=\left(\begin{array}{cccc}
1 & 0 & 0 & 0 \\
0 & \sqrt{1-\eta} & \sqrt{\eta} & 0 \\
0 & -\sqrt{\eta} & \sqrt{1-\eta} & 0 \\
0 & 0 & 0 & 1
\end{array}\right),
\label{eq:unitary}
\end{eqnarray}
where $\eta$ is the interaction parameter controlling the strength of interaction. Notably, the $\eta$ here is identical to the interaction parameter defined in Eq.~\eqref{eq:kraus}. Therefore, if we consider the purified bath in its entirety, the unitary interaction parameter would be of the form: $\tilde{U}^\eta_{PB}=U^\eta_{PB} ~\otimes \mathbb{I}_2$. The channel representation of a thermalization map can also be written in terms of the unitary operator defined above as: $\mathcal{E}(\rho) = \operatorname{Tr}_{bath} U_{PB}(\rho\otimes\rho_{bath})U^{\dag}_{PB}$.

In a more convoluted model, thermalization can be visualized by successive unitary interactions of the probe with many independent purified bath subsystems. Though outside the scope of this work, such a \emph{collisional} model is useful to study the pre-thermalization regime, where the interaction parameter is strictly less than one i.e. $\eta<1$, but the number of interactions can be multiple. 

It is useful to note that the two formulations  -- unitary and Kraus operator -- are equivalent in typical settings i.e. each unitary dilation corresponds to a unique Kraus decomposition and (up to a base change of the initial state and final measurement of the bath) vice versa.  However, when adding quantum control of the external degrees of freedom, different unitary dilations may become physically distinguishable as in the case of Mach-Zehnder interferometer \cite{Wood2021}. We have used each of the aforesaid formalisms for different setups -- namely the unitary interaction  in  Mach-Zehnder, and Kraus operators in the quantum switch --  for our estimation task. Nonetheless, regardless of the quantum control the quantum switch \emph{does not} depend on the particular choice of Kraus decomposition or unitary dilation.

Though we have mostly used a qubit as a probe, in the case of a quantum switch we have also used an $n$ dimensional quantum system widely known as a qudit. For such a probe, the Kraus operators for the GADC can be generalized. Here, we introduce the $n$-dimensional Kraus operator for a general $n$-dimensional (input) thermalization channel below:
\begin{eqnarray}
K_i &=& \sqrt{p_i} \bigg[ |i \rangle \langle i | +\sum_j \sqrt{1-\gamma_{ji}}~~|j \rangle \langle j |  \bigg] \nonumber \\ 
~& \forall & i \in \{1,2,\dots ,n \};~j \neq i ~,\nonumber\\
K_{ij} &=& \sqrt{p_i} \sqrt{\gamma_{ij}}~~| i\rangle \langle j |~~\forall j \neq i ~,
\label{eq:nkraus}
\end{eqnarray}
where $p_i = \frac{e^{-\beta E_i}}{ \sum_i e^{-\beta E_i}}$, and
$K_i$, $K_{ij}$ are the Kraus operators. Physically, they represent an $n$-dimensional system undergoing thermalization, whereby the transition probabilities are: $\gamma_{ij}$ for transition from the level $|j\rangle$ to level $|i\rangle$. In this model, the probabilities of excitation and de-excitation between the same pair of levels are assumed to be equal i.e. $\gamma_{ij} = \gamma_{ji}$.

\subsection{Quantum Processes}

It is useful to introduce the quantum process formalism in the backdrop of the quantum switch, which is a particular process exhibiting indefiniteness in the causal order. This indefiniteness -- as a novel effect -- has been attributed to purported advantages in various fields \cite{Capela_2022, Mukhopadhyay,BAN2021127383,PhysRevA.100.052319,PhysRevResearch.2.033292,PhysRevLett.120.120502, Araujo2014, Taddei2020, Renner2022, PhysRevLett.124.190503, PhysRevA.103.032615, Xie2021, Procopio2021, Yanamandra2023, Gao2022, Das2021, Liu2022, Sen2022, Mo2023}.
A quantum process is a supermap from the set of linear operations (completely positive  -- CP -- maps) on an input Hilbert space $\mathcal{L}(\mathcal{H_A})$, to the set of linear operations on an output Hilbert space $\mathcal{L}(\mathcal{H_B})$. A general $n$-partite quantum process, denoted by $\mathcal{W}$, transforms the set of completely positive trace-preserving (CPTP) maps {$\mathcal{A}_1$,$\mathcal{A}_2$,\dots, $\mathcal{A}_n$} to a new CP map $\mathcal{B}$ such that the input and output Hilbert spaces of the channel $\mathcal{B}$ are identified as past and future \cite{Chiribella_2008}. In general, there could be more than one instance of the same channel being plugged into a given n-partite process. In fact, we only consider a process -- quantum switch -- which has a global past and future, as this allows us to encode the parameter to be estimated in some input state defined on the Hilbert space $\mathcal{P}$ (for ``past''). The output state defined on the Hilbert space $\mathcal{F}$ (for ``future'') is used for measurement and subsequently in estimation.

In our scheme, the channel $\mathcal{B}$ is formed by plugging thermalizing channels corresponding to two distinct temperatures as $\mathcal{A}_i$ into the supermap $\mathcal{W}$. It is also useful to note here that supermaps are insensitive to the choice of dilation (or, equivalently, of Kraus decomposition) of the maps they act on.

The action of the supermap $\mathcal{W}$ is represented by an operator $\mathbb{W}$ which is known as the \textit{process matrix}. This construct allows us to write the Choi representation of the channel $\mathcal{B}$ as follows \cite{Araujo2017}:
\begin{eqnarray} \label{eq:processcomposition}
\mathcal{J}_\mathcal{B} = \operatorname{Tr}_{(i=1,2,\dots,k)} \left[\mathbb{W}^\textrm{\textbf{T}} \left(\otimes^{(i=1,2,\dots,k)}\mathcal{J}_i\otimes \mathbb{I}_\mathcal{B}\right)\right]
\end{eqnarray}
where $\mathcal{J}_\mathcal{B}$ is the Choi representation of the channel $\mathcal{B}$ and $\mathcal{J}_i$ are the Choi representation of channels $\mathcal{A}_i $ and $^\textrm{\textbf{T}}$ represents partial transposition with respect to $\{\mathcal{A}_i \}$. Motivated by physicality of a possible quantum process, the process matrix $\mathbb{W}$ should satisfy certain conditions (see \cite{Araujo_2015} for details). We exemplify in Fig.~\eqref{fig:process} a bipartite quantum process with the channels $\mathcal{A}_1$, $\mathcal{A}_2$ and the resultant channel $\mathcal{B}$, with the same input and output Hilbert spaces -- denoted by $\mathcal{P}$, and $\mathcal{F}$ respectively. 

To summarize, quantum processes allow for more general constructs involving the application of quantum channels and as such, can result in non trivial effects. In our case, we utilize a particular  process called the quantum switch with two distinct thermalizing maps, using the resultant channel ($\mathcal{B}$) for encoding the temperatures on the input state. The output density matrix from the resultant channel is then used to estimate the bounds in Eq.~\eqref{eq:bnds} using Eq.~\eqref{eq:QNM} to calculate the QFIM.

\section{Various sensing setups and bounds on two-temperature estimation} \label{sec:sense}
In this section, we provide a bound on the variances of $T_1$, and $T_2$ and covariances between them using Eq.~\eqref{eq:bnds}. We consider the Mach-Zehnder based setup, and the quantum switch -- incorporating two thermalization channels within these frameworks -- to encode the information about two temperatures in an initialized probe state. The paramaterized state, so obtained, is used to evaluate the bounds. In all of these cases, we only assume a single instance of the experiment, setting $N = 1$. In effect, we calculate the QFIM, and subsequently the bounds and later compare the results based on the range of variances of temperatures obtained for various setups. We also check for the attainability of the bound using Eq.~\eqref{eq:attain}, which is not automatically guaranteed in the multiparamater estimation scenario. Moreover, we fix the input state of the probe and the quantum control. Therefore, the primary aim of this exercise is to highlight the role, and possible validation of the novel probe interaction setups that we have introduced, which essentially rely on coherent quantum control.

\begin{figure}
    \centering
    \includegraphics[width=\columnwidth]{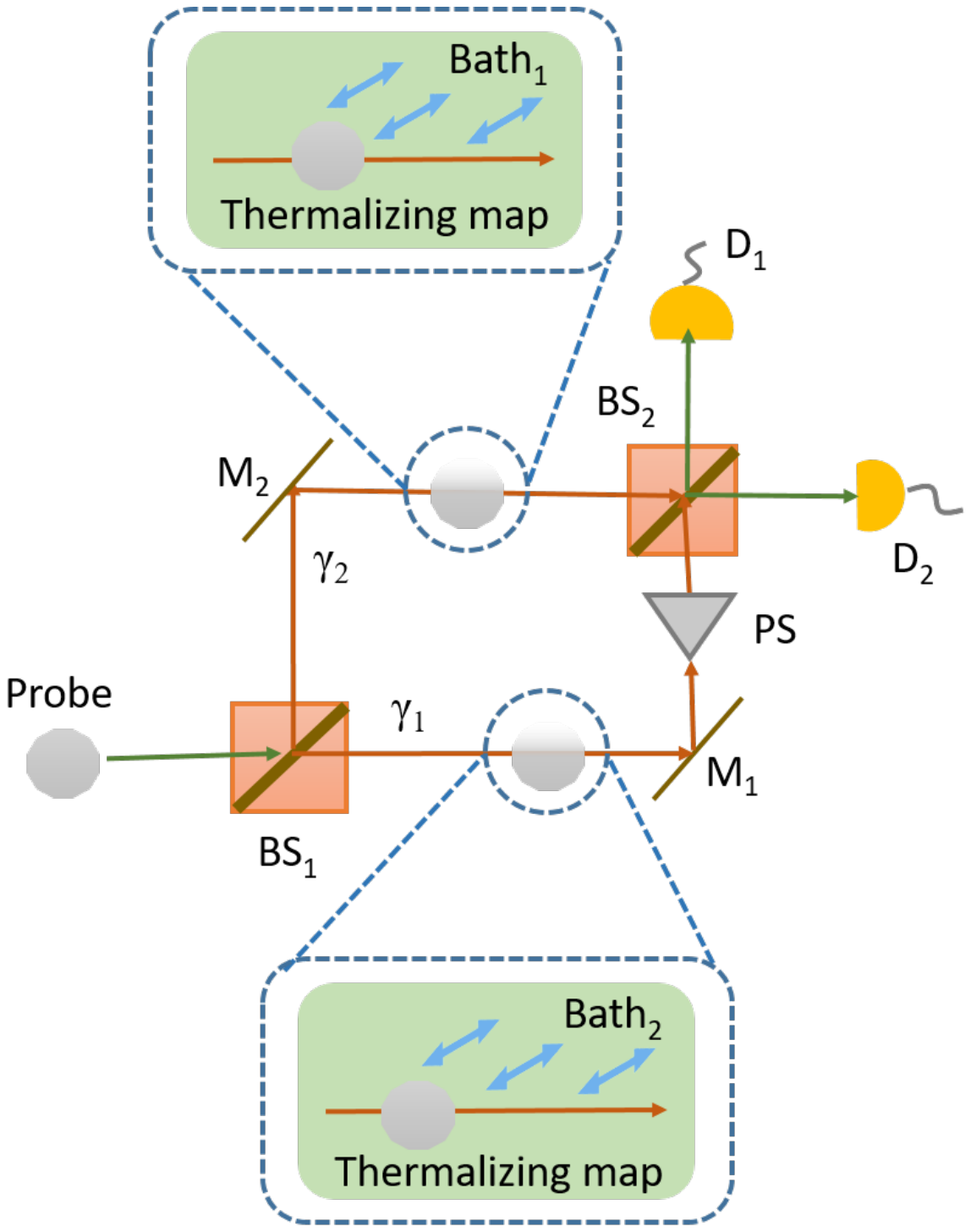}
    \caption{Schematic diagram of the Mach-Zehnder setup -- particularly the two bath case -- to estimate the temperature of the bath(s): Bath$_i$. In the two-bath case, there are two separate baths on each path interacting independently with the probe. $BS_i$, $D_i$ denote the beamsplitters and detectors which act on the path (external DoF) of the thermometer. In one-bath case -- not shown here -- the state of the bath (and its effective temperature) is entangled with the path of the probe (control)}
    \label{fig:MZ}
\end{figure}
\subsection{MZ setup}
We consider a composite particle as a probe with external and internal DoF. In this model of a probe, the internal DoF acts as a thermometer, while  its external DoF -- namely position -- acts as a control and allows the probe to be prepared in a spatial superposition using a beamsplitter. This is shown in Fig.~\eqref{fig:MZ}. In this discussion, we use the terms path and control interchangeably. 

Therefore, if the probe interacts with baths at different temperatures, or a single bath with its temperature conditioned on the control, the information about both the temperatures gets encoded in the internal state of the probe, if the control is prepared in a spatial superposition. This scheme was introduced in \cite{Wood2021}, while considering a fundamental question of assigning quantum superposition to an essential thermodynamic quantity: temperature. Here, we lay a particular emphasis on adopting the methodology introduced therein for a particular application: two-temperature thermometry.

We consider here the thermometer to be a qubit that is initially prepared in the internal state $|\psi_0\rangle$. Subsequently it is prepared in a spatial superposition, after moving through a beamsplitter -- $BS_1$. At this stage, the state of the probe can be written as a product state of its internal and external DoF as follows:
\begin{eqnarray}
|\psi\rangle_{BS_1}=|\psi_0\rangle \otimes \frac{ (|\gamma_1\rangle + |\gamma_2\rangle)}{\sqrt{2}} \equiv \frac{1}{\sqrt{2}}\big[|\psi_0\rangle|\gamma_1\rangle + |\psi_0\rangle|\gamma_2\rangle\big], \nonumber\\
\label{eq:1}
\end{eqnarray}
where $|\gamma_1\rangle \leftrightarrow \gamma_1$, and $|\gamma_2\rangle \leftrightarrow \gamma_2$ refer to the path DoF.

As the probe moves through the arms of the interferometer, it interacts with the thermal bath. One of the arms ($\gamma_1$) has a phase-shifter (PS) which allows for an external control over the relative phase between the paths. The baths are considered to be purified thermal states, as in Eq.~\eqref{eq:bath}. The interaction between the bath and the thermometer is governed by the unitary operator in Eq.~\eqref{eq:unitary} acting on the probe and bath DOF. It is important to note that in all of the cases, we have assumed the initial internal state of the probe to be $|\psi_0\rangle = |0\rangle$ though the results are independent of the initial state as long as we consider full thermalisation in each arm. In such a setup, we can imagine two different scenarios as introduced in \cite{Wood2021}, outlined below.
\begin{figure*}
    \centering
    \includegraphics{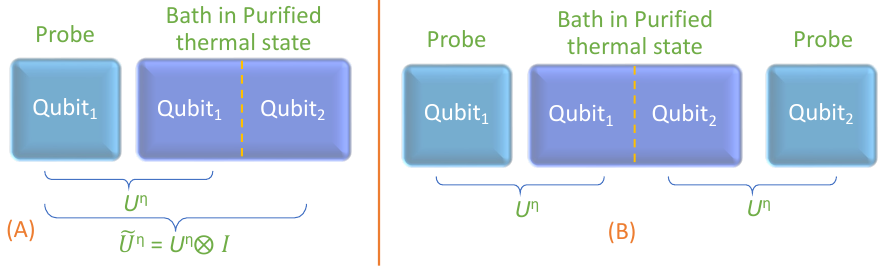}
    \caption{Interaction scheme of single-qubit probe (A) and two-qubit probe (B) used in the Mach-Zehnder setup.}
    \label{fig:mz_2n1}
\end{figure*}
\subsubsection{One-bath case}
\begin{figure}
     \centering
     \begin{subfigure}[b]{0.85\columnwidth}
         \centering
         \includegraphics[width=\columnwidth]{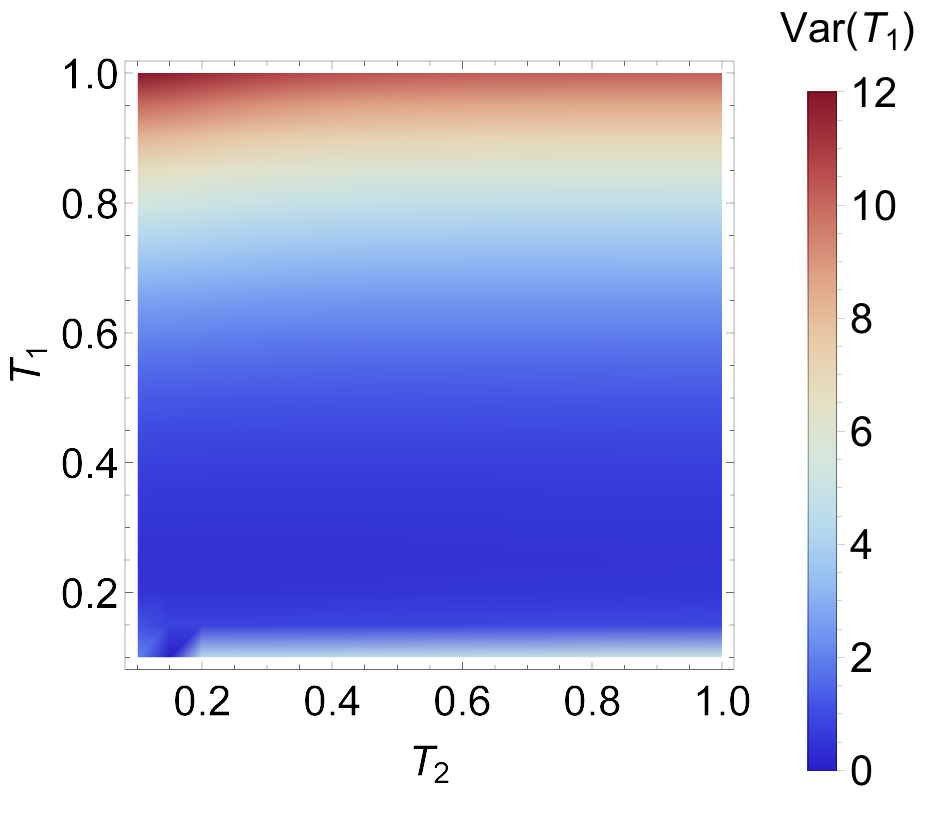}
         \caption{$\operatorname{Var}(T_1)$}
         \label{fig:1b_v1}
     \end{subfigure}
     %\hfill
     \begin{subfigure}[b]{\columnwidth}
         \centering
         \includegraphics[width=\columnwidth]{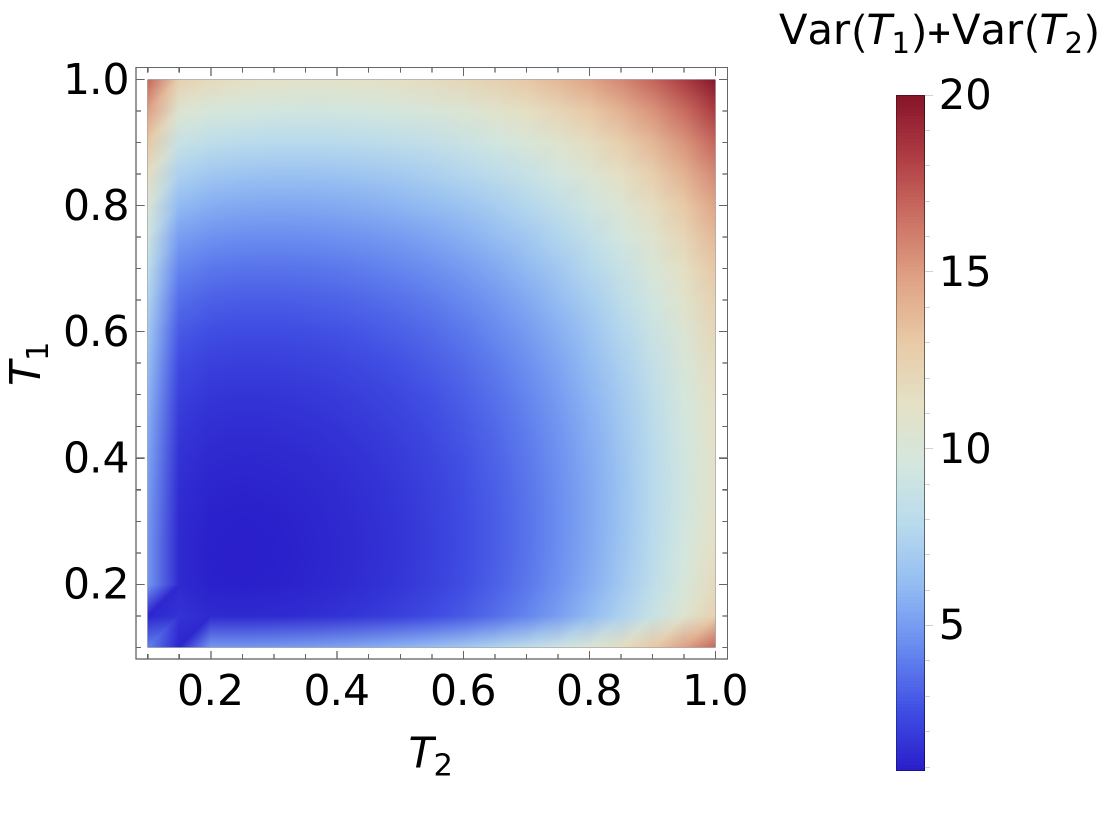}
         \caption{$\operatorname{Var}(T_1)+ \operatorname{Var}(T_2)$}
         \label{fig:1b_v2}
     \end{subfigure}
        \caption{One-bath case with two-qubit probe with its state given in Eq.~\eqref{eq:1b_ps} (a) Heatmap of the variance of the estimated temperature $T_1$ as a function of the temperatures $T_1$ and $T_2$. A similar heatmap of the estimated temperature $T_2$ can be obtained by interchanging the axes' labels due to the symmetric nature of the setup. (b) Heatmap of the total variance. $\phi=\pi/2$.}
        \label{fig:1_bath}
\end{figure}

\begin{figure}
     \centering
     \begin{subfigure}[b]{0.85\columnwidth}
         \centering
         \includegraphics[width=\columnwidth]{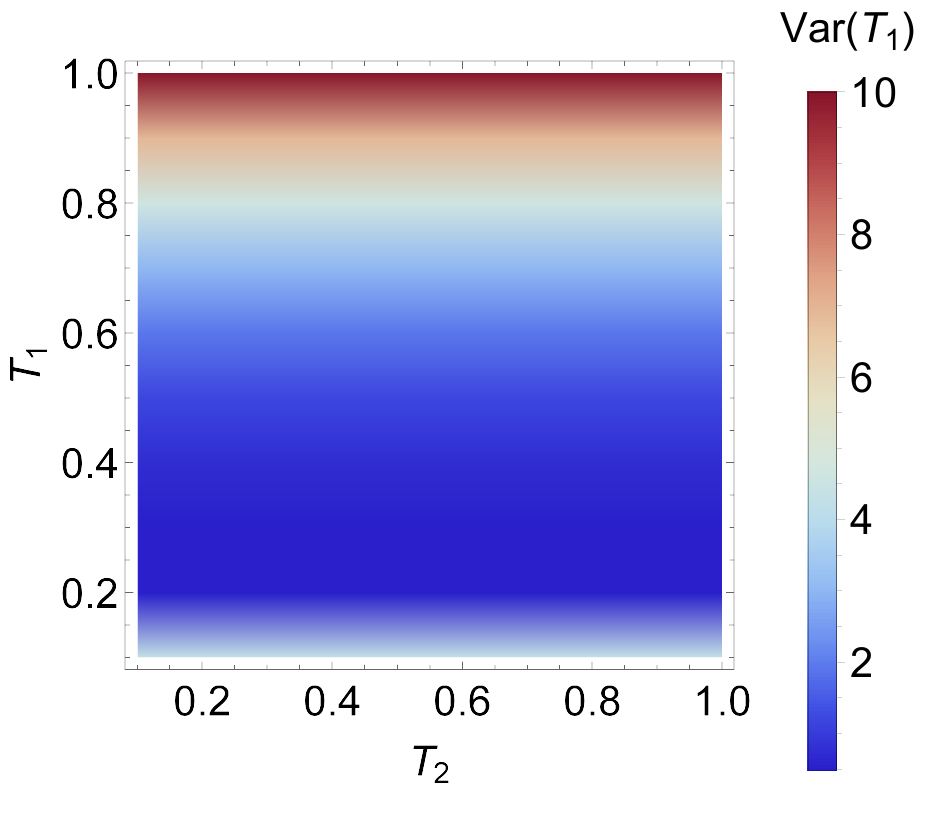}
         \caption{$\operatorname{Var}(T_1)$}
         \label{fig:1b_wc_v1}
     \end{subfigure}
     %\hfill
     \begin{subfigure}[b]{\columnwidth}
         \centering
         \includegraphics[width=\columnwidth]{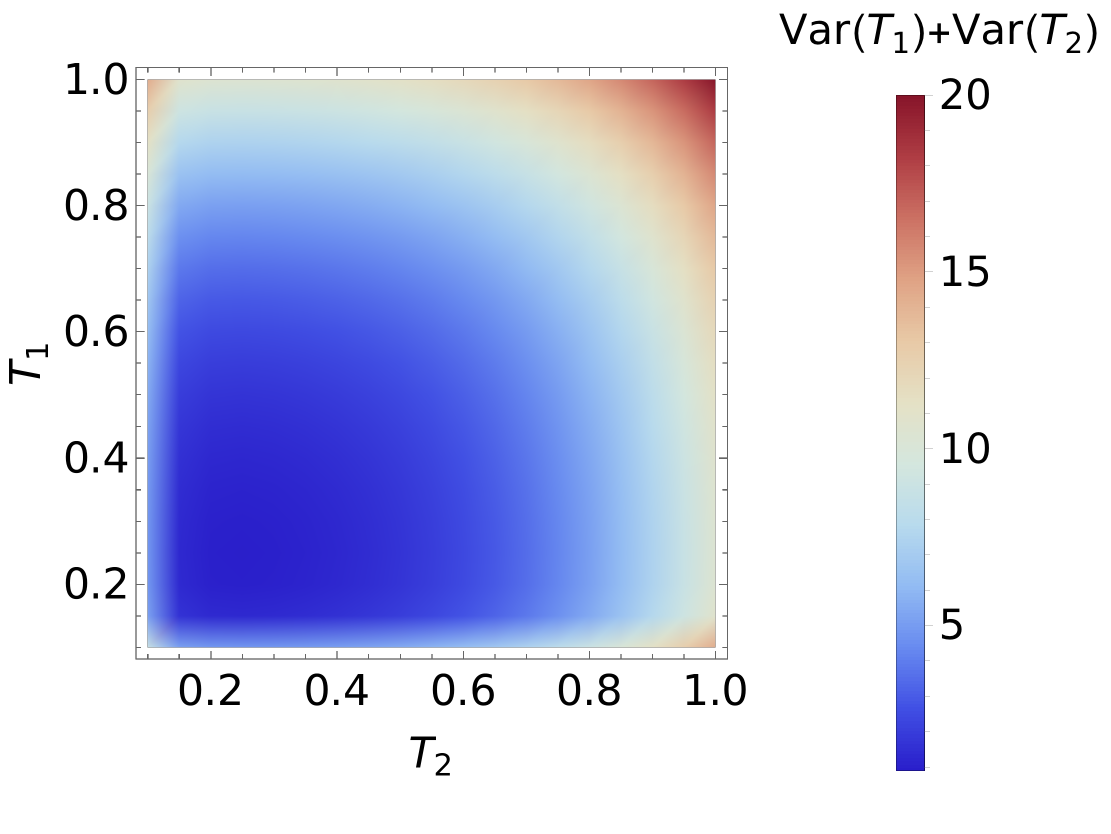}
         \caption{$\operatorname{Var}(T_1)+ \operatorname{Var}(T_2)$}
         \label{fig:1b_wc_v2}
     \end{subfigure}
        \caption{One-bath case with estimation based on the joint state of a single qubit probe and the control, obtained from Eq.~\eqref{eq:1b_full} after tracing out the bath.(a) Heatmap of the variance of the estimated temperature $T_1$ as a function of the temperatures $T_1$ and $T_2$. A similar heatmap of the estimated temperature $T_2$ can obtained by interchanging the axes' labels due to the symmetric nature of the setup. (b) Heatmap of the total variance. $\phi=\pi/2$.}
        \label{fig:1b_wc}
\end{figure}
In this scenario, the (purified) state of the bath -- and by extension its effective temperature -- is contingent on the control i.e the path DoF of the probe, and hence is also quantum-controlled. Therefore, the state of the probe, along with the bath, before it goes through the beamsplitter $BS_2$ can be written as:
\begin{eqnarray}
|\psi\rangle_{BS_2} =  \frac{1}{\sqrt{2}}\big[e^{i\phi}\tilde{U}^\eta_{PB}|\psi_0\rangle |\theta^{\beta_1}\rangle|\gamma_1\rangle +\tilde{U}^\eta_{PB} |\psi_0\rangle|\theta^{\beta_2}\rangle|\gamma_2\rangle\big], \nonumber\\
\label{eq:1b_full}
\end{eqnarray}
where $\tilde{U}^\eta_{PB}$ is from Eq.~\eqref{eq:unitary}.
The beamsplitter $BS_2$ along with the detectors $D_1$, and $D_2$ is used for a measurement of the path DOF of the probe in the basis $|\pm\rangle_\gamma = \frac{1}{\sqrt{2}}[|\gamma_1\rangle \pm |\gamma_2\rangle ]$. It is assumed that the interaction of the probe with the bath is complete and subsequently, the bath is traced out at this stage. Hence, the post-selected non-normalized (internal) state of the probe is given as:
\begin{eqnarray}
\rho_\pm &=& \frac{1}{2}\operatorname{Tr}_B\big[(\tilde{U}^\eta_{PB})|\psi_0\rangle |\theta^{\beta_1}\rangle\langle\theta^{\beta_1}|\langle\psi_0|(\tilde{U}^\eta_{PB})^{\dagger}\nonumber\\
&+&(\tilde{U}^\eta_{PB})|\psi_0\rangle |\theta^{\beta_2}\rangle\langle\theta^{\beta_2}|\langle\psi_0|(\tilde{U}^\eta_{PB})^{\dagger}\nonumber\\
&\pm&e^{i\phi}(\tilde{U}^\eta_{PB})|\psi_0\rangle |\theta^{\beta_1}\rangle\langle\theta^{\beta_2}|\langle\psi_0|(\tilde{U}^\eta_{PB})^{\dagger}\nonumber\\
&\pm&e^{-i\phi}(\tilde{U}^\eta_{PB})|\psi_0\rangle |\theta^{\beta_2}\rangle\langle\theta^{\beta_1}|\langle\psi_0|(\tilde{U}^\eta_{PB})^{\dagger}\big].
\label{eq:1b_ps}
\end{eqnarray}
Post normalization, the $\rho_+$ state is used in the estimation task of the two temperatures: here represented by their inverses: $\beta_1$, and $\beta_2$.

Through numerical modelling of the bath and its interaction with the probe, we obtain the paramaterized state of the probe, which carries information about the two distinct temperatures. Using this parameterized state to calculate the  QFIM, we calculate the bounds and arrive at the following key results:
\begin{itemize}
    \item \textbf{Single-qubit probe, with the control measured in the $\pm$ basis:} For a single qubit probe with the postselected state in Eq.~\eqref{eq:1b_ps}, the determinant of the QFIM is \textbf{zero} for all values of $\phi$. Therefore, it follows that a single qubit cannot be used for measuring two distinct temperatures in such a setup. Note that in this case, one could not have made such a conclusion intuitively.
    
    \item \textbf{Single-qubit probe, with arbitrary probe-control measurement:} A single qubit probe, used in conjugation with the control, such that the estimation is now performed on the joint state of the control and probe. The relevant state in this case is obtained from Eq.~\eqref{eq:1b_full} after tracing out the bath. We find that the QFIM is non-singular, and therefore, evaluate the bounds in  Eq.~\eqref{eq:bnds}. We show the variances and covariances of temperatures in Fig.~\eqref{fig:1b_wc}. We also find that the attainability criteria is satisfied for the complete range of temperatures considered. Our calculations indicate that the variances in this case are independent of $\phi$.
    
    \item \textbf{Two-qubit probe:} We consider a scenario where two qubits are used as a probe, with the unitary interaction modified in a way that each of these qubits interacts with one of the qubits of the purified bath through $U_{PB}^\eta$. This scheme is shown in Fig.~\eqref{fig:mz_2n1}. We find that the determinant of QFIM is finite in this case, with some caveats given below. In the case where the QFIM is non-singular ( with $\phi = \pi/2$), we calculate the bounds in Eq.~\eqref{eq:bnds}, and we show the dependencies of variance and covariance of temperatures in Fig.~\eqref{fig:1_bath}.
    It is noteworthy that $\phi$ also controls the nature of the QFIM. For some particular values of $\phi$, the QFIM is singular, indicating that the estimation task cannot be accomplished, e.g. if $\phi=0$. In order to show that that the simultaneous estimation task can, in principle be achieved, we have selected the phase induced by $PS$ to be $\phi=\pi/2$ in Fig.~\eqref{fig:1_bath}. We also observe that in this particular case, the attainability criteria in Eq.~\eqref{eq:attain} is satisfied, meaning that the bound can indeed be saturated.
\end{itemize}

On comparing the cases of estimation task using the joint state of a single qubit and the control, with that of two qubits while the control is measured, we find that the former performs slightly better. This is inferred using the maxima of variances from Fig.~\eqref{fig:1_bath} and Fig.~\eqref{fig:1b_wc}.

\subsubsection{Two-bath case}
\begin{figure}
     \centering
     \begin{subfigure}[b]{0.85\columnwidth}
         \centering
         \includegraphics[width=\columnwidth]{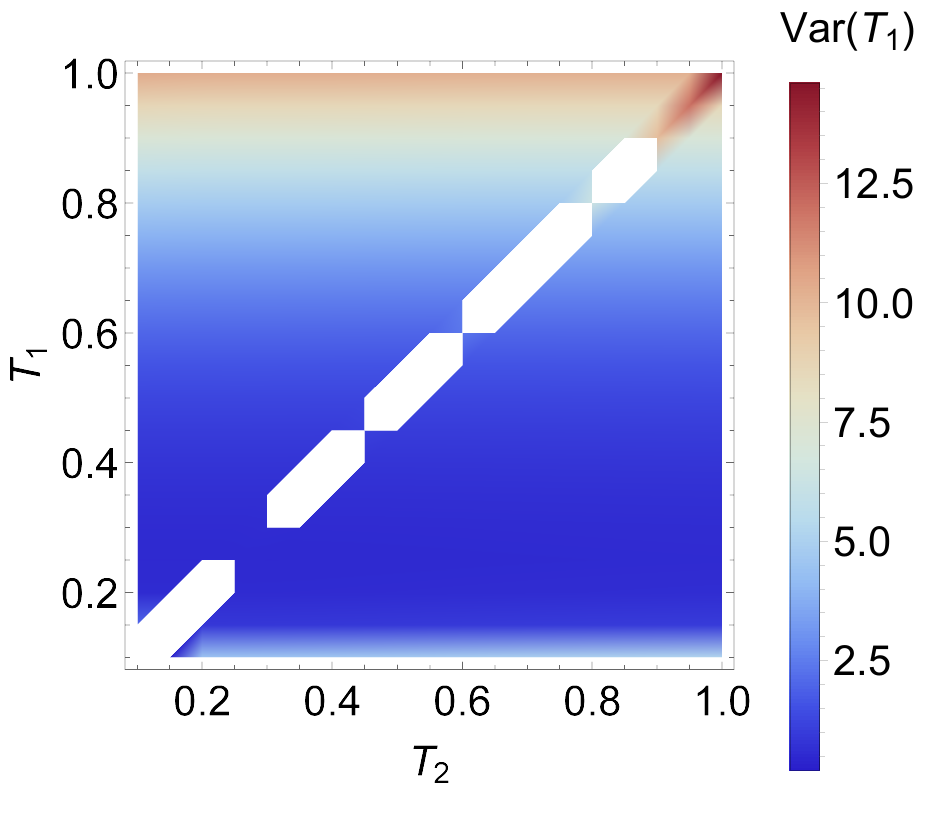}
         \caption{$\operatorname{Var}(T_1)$}
         \label{fig:2b_v1}
     \end{subfigure}
     %\hfill
     \begin{subfigure}[b]{\columnwidth}
         \centering
         \includegraphics[width=\columnwidth]{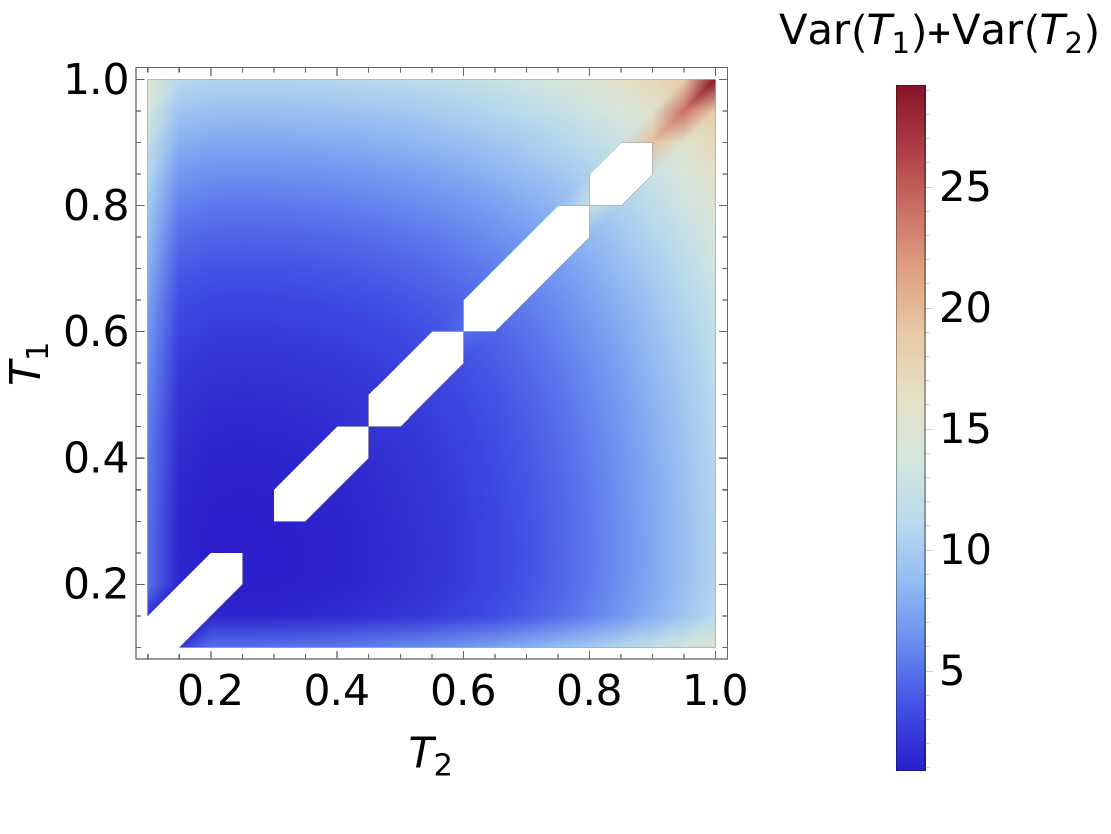}
         \caption{$\operatorname{Var}(T_1)+ \operatorname{Var}(T_2)$}
         \label{fig:2b_v2}
     \end{subfigure}
        \caption{Two-bath case with two qubit probe state given in Eq.~\eqref{eq:rhoout}.(a) Heatmap of the variance of the estimated temperature $T_1$ as a function of the temperatures $T_1$ and $T_2$. A similar heatmap of the estimated temperature $T_2$ can obtained by interchanging the axes' labels due to the symmetric nature of the setup. (b) Heatmap of the total variance. The white parts denote areas where the variance diverges. $\phi=\pi/2$.}
        \label{fig:2_bath}
\end{figure}
\begin{figure}
     \centering
     \begin{subfigure}[b]{0.85\columnwidth}
         \centering
         \includegraphics[width=\columnwidth]{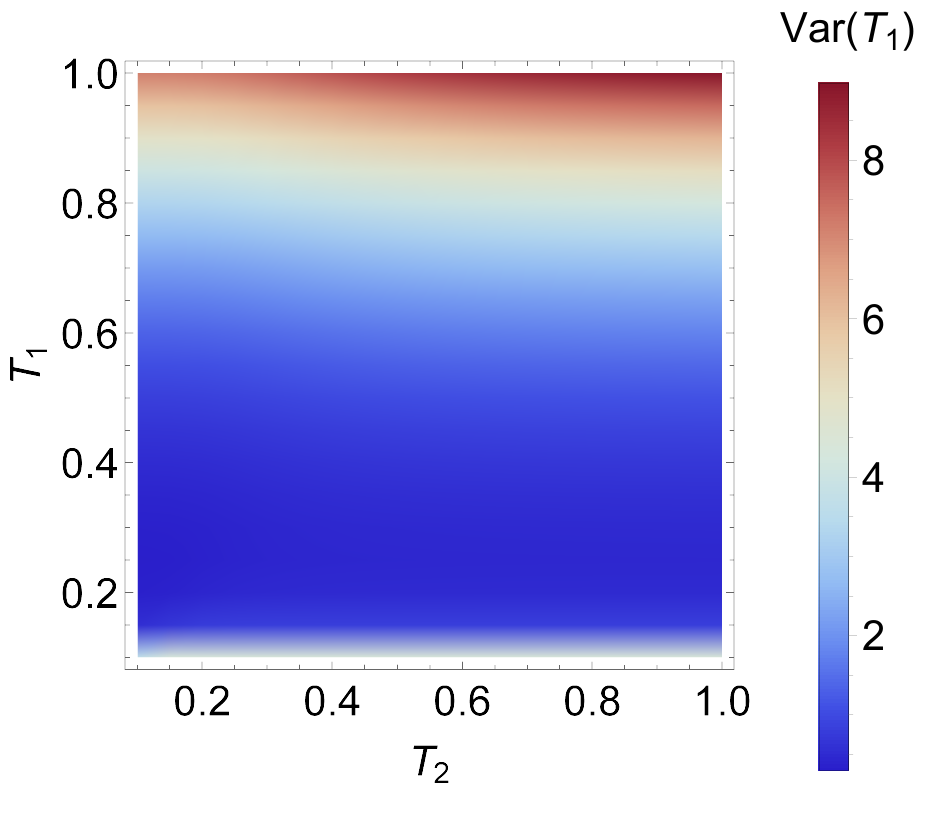}
         \caption{$\operatorname{Var}(T_1)$}
         \label{fig:2b_wc_v1}
     \end{subfigure}
     %\hfill
     \begin{subfigure}[b]{\columnwidth}
         \centering
         \includegraphics[width=\columnwidth]{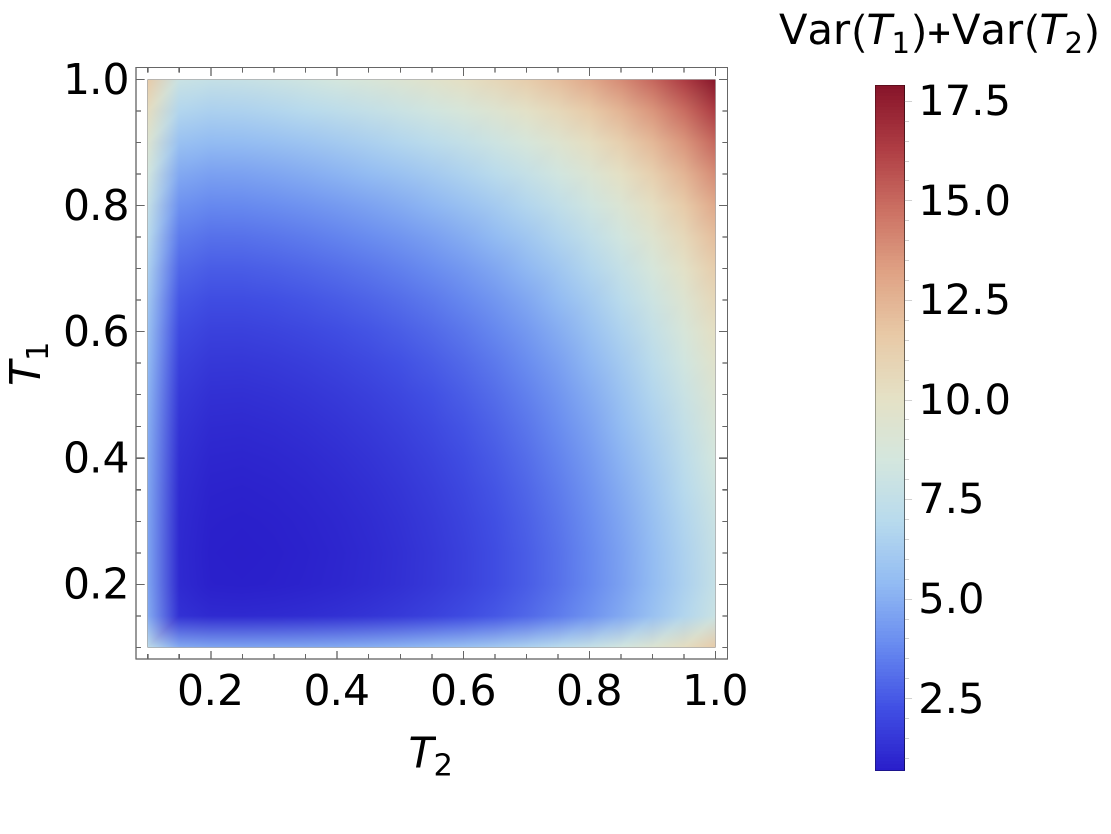}
         \caption{$\operatorname{Var}(T_1)+ \operatorname{Var}(T_2)$}
         \label{fig:2b_wc_v2}
     \end{subfigure}
        \caption{Two-bath case with estimation based on the joint state of a single qubit probe and the control, obtained from Eq.~\eqref{eq:2} after tracing out the bath. Notably, the exact same variances are obtained for the quantum switch setup.(a) Heatmap of the variance of the estimated temperature $T_1$ as a function of the temperatures $T_1$ and $T_2$. A similar heatmap of the estimated temperature $T_2$ can obtained by interchanging the axes' labels due to the symmetric nature of the setup. (b) Heatmap of the total variance. $\phi=\pi/2$.
        }
        \label{fig:2b_wc}
\end{figure}
In this scenario, two independent baths are considered along the paths $\gamma_1$ and $\gamma_2$. Considering the baths on the paths $\gamma_1$ and $\gamma_2$ are in the states $|\theta^{\beta_1}\rangle$ and $|\theta^{\beta_2}\rangle$ respectively, the initial state of the complete system i.e. the baths and the probe after going through the beamsplitter $BS_1$ can be written as follows:
\begin{eqnarray}
|\psi\rangle_{BS_1}=\frac{1}{\sqrt{2}}\underbracket{|\psi_0\rangle}_{Internal~DoF} \otimes \underbracket{(|\gamma_1\rangle + |\gamma_2\rangle)}_{Path~DoF} \otimes \underbracket{|\theta ^{\beta_1}\rangle \otimes |\theta^{\beta_2}\rangle}_{Bath~DoF}~.\nonumber\\
\end{eqnarray}
The probe then undergoes a unitary interaction with the baths such that the state of the complete system (including the baths) after evolving through the arms of the interferometer and just before $BS_2$ is:
\begin{eqnarray}
|\psi\rangle_{BS_2} = \frac{1}{\sqrt{2}}\big[ e^{i\phi}|\gamma_1\rangle U_{\gamma_1} |\psi_0\rangle |B\rangle +|\gamma_2\rangle  U_{\gamma_2} |\psi_0\rangle|B\rangle\big]~,\nonumber \\
\label{eq:2}
\end{eqnarray}
where $|B\rangle = |\theta^{\beta_1}\rangle \otimes |\theta^{\beta_2}\rangle$, $U_{\gamma_1} = \tilde{U}^\eta_{PB_1}\otimes \mathbb{I}_{B_2}$, and $U_{\gamma_2}=\tilde{U}^\eta_{PB_2}\otimes \mathbb{I}_{B_1}$. This implies that the unitary interaction in each arm is between the probe and the (independent) bath specific to the particular path.

Similar to the one-bath case, if the path of the probe is measured in the diagonal basis $|\pm\rangle_\gamma$, and the bath qubits are traced out, the following (non-normalized) density matrix of the internal DoF of the probe is obtained:
\begin{eqnarray}
\rho_\pm&=&  \frac{1}{2}\operatorname{Tr}_B\big[U_{\gamma_1} (|\psi_0\rangle \otimes |B\rangle)(\langle\psi_0| \otimes \langle B|)U_{\gamma_1}^{\dagger}\nonumber\\
&+& U_{\gamma_2} (|\psi_0\rangle \otimes |B\rangle)(\langle\psi_0| \otimes \langle B|)U_{\gamma_2}^{\dagger}\nonumber\\
&\pm&e^{i\phi} U_{\gamma_1} (|\psi_0\rangle \otimes |B\rangle)(\langle\psi_0| \otimes \langle B|)U_{\gamma_2}^{\dagger}\nonumber\\
&\pm&e^{-i\phi}U_{\gamma_2} (|\psi_0\rangle \otimes |B\rangle)(\langle\psi_0| \otimes \langle B|)U_{\gamma_1}^{\dagger} \big].
\label{eq:rhoout}
\end{eqnarray}
We use the above state for the simultaneous temperature estimation task, after normalization. The key results that we find are listed below.
\begin{itemize}
    \item \textbf{Single-qubit probe, with the control measured in the $\pm$ basis:} Similar to one-bath case, we find that the QFIM obtained is singular for all values of $\phi$. Therefore, it cannot be used to provide any bound on the variance.
    
    \item \textbf{Single-qubit probe, with arbitrary probe-control measurement:} If the single qubit is used as a probe in conjugation with the control i.e. the state in Eq.~\eqref{eq:2} after tracing out the baths, we find that the QFIM is non-singular, and subsequently, the bound in Eq.~\eqref{eq:bnds} can be evaluated. We show the covariance and the variances of temperature in Fig.~\eqref{fig:2b_wc}. We also find that the attainability criteria is satisfied for the complete range of temperatures considered here. Moreover, our calculations show that the variances obatined are not dependent on $\phi$ in this case.
    
    \item \textbf{Two-qubit probe:} The unitary interaction between the probe and the bath is of the form: $U_{\gamma_1} = U_{PB}^\eta \otimes U_{PB}^\eta \otimes \mathbb{I}_{B_2}$, and similarly for $U_{\gamma_2}$, where the $U_{PB}^\eta$ are qubit-pairwise interactions between the probe and the bath (nature of interaction shown in Fig.~\eqref{fig:mz_2n1}). It is seen here that in certain instances, where $T_1=T_2$, there may be some non-finite terms while evaluating the bounds in Eq.~\eqref{eq:bnds}, which is shown in Fig.~\eqref{fig:2_bath}. In other instances, it is established that the variances are finite and hence, the simultaneous estimation task can be accomplished, in principle. In most of the temperature regime shown here, the attainability criteria is tested and found to be satisfied. However, for some temperatures, this is not true, meaning that the bounds obtained cannot be saturated.
    Similar to the one-bath case, the QFIM here is also susceptible to the phase introduced by $PS$ in the arm $\gamma_1$. We have uniformly set $\phi=\pi/2$ to demonstrate the success of the simultaneous estimation task. However, for $\phi=0$, we get a singular QFIM and hence, the above bounds do not exist.
\end{itemize}

On comparing the results from Fig.~\eqref{fig:2_bath} and Fig.~\eqref{fig:2b_wc}, we see that the case with the joint state of a single qubit probe and the control fares better than the 2 qubit probe, with the control measured in a coherent basis. Moreover, in this case the variances do not diverge at any temperature, and the attainability criteria is always satisfied.

\subsection{Quantum Switch}
There have been various proposals demonstrating advantage in tasks related to metrology \cite{Mukhopadhyay,BAN2021127383}, communication \cite{PhysRevA.100.052319,PhysRevResearch.2.033292,PhysRevLett.120.120502}, and computation \cite{Araujo2014, Taddei2020, Renner2022} using the quantum switch. The switch sits within the broader framework of quantum processes, which includes processes exhibiting indefinite causal order \cite{Oreshkov2012}, and time-ordered processes such as quantum combs \cite{Chiribella2009}. There have been recent efforts to incorporate quantum processes in parameter estimation and channel discrimination tasks \cite{PhysRevLett.124.190503, PhysRevA.103.032615, Xie2021}.

\begin{figure}[b]
    \centering
    \includegraphics[scale = 0.85]{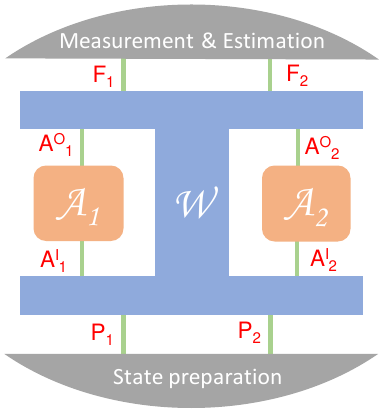}
    \caption{Schematic diagram of a bipartite quantum switch with all the Hilbert spaces marked. The input state is prepared as: $ |0\rangle \langle 0|_{P_1}$ (target) $\otimes |+\rangle\langle +|_{P_2}$ (control). This control state leads to a superposition in the causal order, specifically in the order of application of the maps $\mathcal{A}_1$, and $\mathcal{A}_2$ on the target. The estimation task is performed after measurement on the joint output state defined on the Hilbert space $F_1 \otimes F_2$ i.e. on target+control.}
    \label{fig:swi}
\end{figure}
The quantum switch is a quantum process that allows quantum control over the order of application of two or more channels. For example, if we consider two quantum maps $\mathcal{A}_1$, and $\mathcal{A}_2$, the quantum control can be exerted over the compositions: $\mathcal{A}_1\circ\mathcal{A}_2$ and $\mathcal{A}_2\circ\mathcal{A}_1$. The resultant channel $\mathcal{B}$ takes as input two states: $\rho_\textrm{in}$ (also called target), and $\rho_A$ (also called ancilla or control), and outputs state: $\rho_\textrm{out}$. If the control is prepared in the state $|0\rangle$, then the output is: $\rho_\textrm{out}= \mathcal{A}_1\circ\mathcal{A}_2 (\rho_\textrm{in})$. Similarly for the control input $|1\rangle$, the output is:  $\rho_\textrm{out}=\mathcal{A}_2\circ\mathcal{A}_1 (\rho_\textrm{in})$. The preparation of the control as a probablisitic mixture leads to a probablistic application of the composition of $\mathcal{A}~s$ on the input state. However, if the control is prepared in the state $\rho_A= |+\rangle = \frac{1}{\sqrt{2}} [|0\rangle+|1\rangle$], the application of an equal superposition of the composition of the channels on $\rho_\textrm{in}$ is realized. We state this effect in the Kraus representation below. Unlike the MZ setup, the switch is insensitive to the choice of dilation of the channels (equivalently, it is independent of the choice of Kraus representation).

A bipartite quantum switch, with all its input and output Hilbert spaces marked is denoted in Fig.~\eqref{fig:swi}. On the input side, the target is defined on the Hilbert space $P_1$, and the control on $P_2$. Also, on the output side, $F_1$ and $F_2$ can be identified as the Hilbert spaces of the target and control respectively. The action of switch on the input i.e. the probe and control can be simply written using the Kraus representation of the individual channels, $\mathcal{A}_1 (\rho) = \sum_i F_i \rho F_i^\dagger$, and $\mathcal{A}_2 (\rho) = \sum_j K_j \rho K_j^\dagger$, as follows:
\begin{align}
    \mathcal{B} (\rho_{SC}) &= \sum_{ij} H_{ij}(\rho_{SC}) H_{ij}^\dagger~,\nonumber\\
    H_{ij} &= F_i K_j \otimes |0\rangle \langle 0|_C + K_j F_i \otimes |1\rangle \langle 1|_C~,
\end{align}
where the subscript $_C$ denotes the control. Also $\mathcal{B}$ is the resultant channel obtained using the switch with $\mathcal{A}_1. \mathcal{A}_2$, and $\rho_{SC}$ is defined as the joint input state on target and control. It is noteworthy the process belongs to a category of quantum processes with indefinite causal order, if the control is prepared in a superposition. 

In the process matrix formalism, a bipartite quantum switch is represented as follows \cite{Araujo_2015}:
\begin{eqnarray}
\mathbb{W}_\textrm{switch}&=&|w_\textrm{switch}\rangle\langle w_\textrm{switch}|~,\nonumber\\
|w_{\textrm{switch }}\rangle &=&|0\rangle^{P_{1}}|\mathbbm{1}\rangle\rangle^{P_{2} A^{I}_1}|\mathbbm{1}\rangle\rangle^{A_1^{O} A_2^{I}}|\mathbbm{1}\rangle\rangle^{A_2^{O} F_{2}}|0\rangle^{F_{1}}\nonumber\\
&+&|1\rangle^{P_{1}}|\mathbbm{1}\rangle\rangle^{P_{2} A_2^{I}}|\mathbbm{1}\rangle\rangle^{A_2^{O} A_1^{I}}|\mathbbm{1}\rangle\rangle^{A_1^{O} F_{2}}|1\rangle^{F_{1}}~,\nonumber\\
\end{eqnarray}
where $|\mathbbm{1}\rangle\rangle =\sum_j |j\rangle |j\rangle$, with  $\{|j\rangle\}_j$ as orthonormal basis, represents the Choi form of the identity map.
We use Eq.\eqref{eq:processcomposition} with the GADC channel corresponding the distinct temperatures, and the process matrix for the quantum switch to obtain the resultant channel. For a qubit as the target, we plug in the state: $\rho_\textrm{in} = |\psi_0\rangle\langle \psi_0| \otimes  |+\rangle\langle +|$ as the target + control. The joint state of the target and control is then used to estimate the temperatures through a joint measurement.

We find that the QFIM is non-singular in this case. Hence, two temperatures can be measured simultaneously using this setup. The behaviour of the variance is qualitatively and quantitatively identical to that of the two-bath case with the estimation performed on the probe + control DoFs. Therefore, the same results follow from Fig.~\eqref{fig:2b_wc}. 

Moreover, we also consider an n-dimensional system -- qudit -- with $n=3,~4$ as the target along with qubit as the control in the quantum switch, with the appropriate thermalization channel (from Eq.~\eqref{eq:nkraus}) and a modified process matrix of the switch admitting such a target. We find that in all of these cases with a switch based setup, the attainability condition is satisfied. We calculate the range of variances offered by such n-dimensional setups and show them in Fig.~\eqref{fig:comp} for a comparison with other setups.

\subsection{Comparing various setups}
\begin{figure}
    \centering
    \includegraphics[width=\columnwidth]{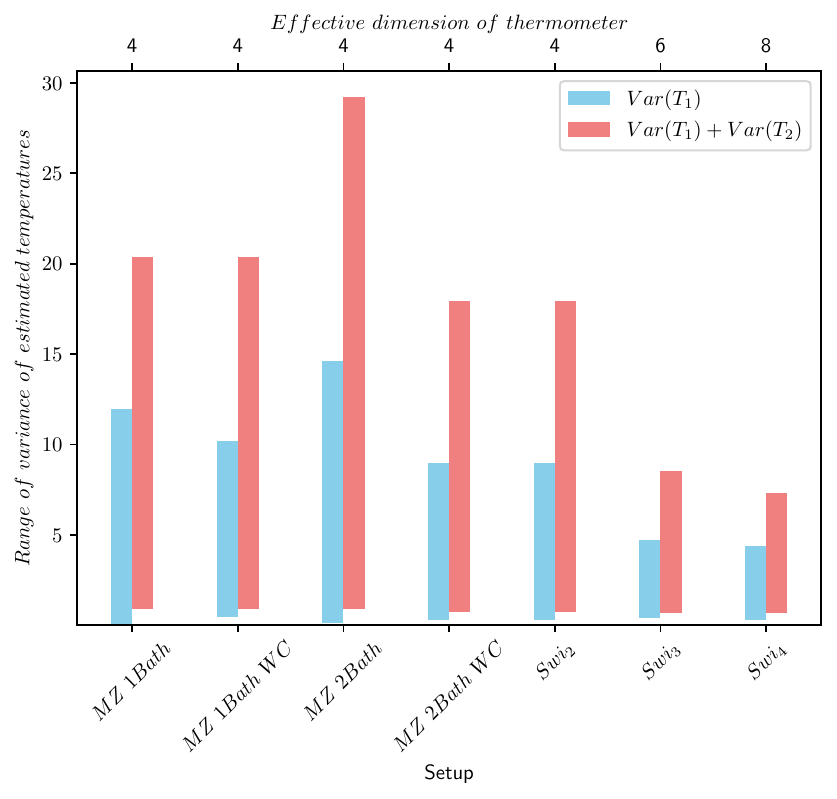}
    \caption{Range of the variance of estimated temperature (Blue), alongside the total variance (Red) obtained for the setups discussed in Sec.~\ref{sec:sense}. The input state of the probe is considered to be $|0\rangle \langle 0|$ in all cases, while setting $N =1$. We indicate $Swi_n \rightarrow $ switch setup with an n-dimensional target, and a qubit as control, and the addendum $WC$ for Mach-Zehnder setups with the control included in the estimation task. The maximum variance and the total variance offered by $Swi_3$, $Swi_4$ is much less than that offered by Mach-Zehnder based setups and $Swi_2$. This shows that an advantageous strategy in the two-temperature estimation can be envisaged by increasing the dimensionality of the probe in a switch based setup. Since the MZ two-bath case has some infinite terms in the variance (refer Fig.~\eqref{fig:2_bath}), the range has been calculated after ignoring these infinite terms. The effective dimension of the thermometer -- indicated in the second x-axis -- represents the dimension of the probe and the control considered together in the relevant cases where their joint state is used in the estimation task.}
    \label{fig:comp}
\end{figure}

In the quantum switch based setup that we have considered, the measurement and estimation task is assumed to be performed on the complete output space associated with the process -- i.e. including the quantum control. It is important to consider the effective dimensionality of the thermometric setup -- which includes quantum control in certain instances -- in making a fair comparison between various setups. It is clear from the results illustrated before that in all the instances, the variance of one temperature (say $T_1$) is a function of both the temperatures ($T_1$ and $T_2$), though it is largely dependent on the temperature to be estimated ($T_1$ in this example). It is a priori unclear how to compare the variances over the complete parameter space spanned by $T_1$, and $T_2$. Therefore, we resort to defining the range of variance i.e. the minimum and maximum variance offered by each setup for the temperature $T_1$ as well as the range of total variance $\operatorname{Var}(T_1)+ \operatorname{Var}(T_2)$.

In Fig.~\eqref{fig:comp}, we have shown these ranges through a bar plot spanning from the minimum variance to the maximum variance. We note from the figure that if the dimensionality of the target is increased in the switch, for example in the setups denoted by $Swi_3$, and $Swi_4$, the range of variance is further limited, indicating an advantage. However, such an advantage comes with at an additional requirement of a higher dimensional thermometer which cannot be ignored while making a fair comparison with lower dimensional setups. Nevertheless, such  multidimensional thermometers may find applications in high-precision two-temperature thermometry, where the dimensionality aspect can be overlooked in favor of better performance. If we resort to limiting our study to a qubit probe -- i.e. the cases of MZ 1-bath, MZ 1-bath with control, MZ 2-bath, MZ 2-bath with control and $Swi_2$ -- for a fair comparison, we find that the MZ 2-bath with control performs equally well as the $Swi_2$ setup (see Fig.~\eqref{fig:2b_wc}) and both of these setups outdo others in a majority of parameter space of $T_1 \in [0.1,1]$, and $T_2 \in [0.1,1]$. However, it is noteworthy that the MZ 1-bath setup performs better than the above in the low temperature regime. This is also reflected in Fig.~\eqref{fig:comp}, where range of the MZ 1-bath setup goes lower than that of others of the same dimensionality of the probe.

\section{Conclusions} \label{sec:conc}

To recall, the QFIM provides a bound on the covariance matrix, whose diagonal and off-diagonal elements are the variances and the covariances of (between) the two temperatures respectively. It must be reiterated that such a bound may not exist if the QFIM is singular and that the bound may not be attainable even if the QFIM is non-singular. Apart from this, the laboratory conditions may restrict the class of possible operations -- measurements and hence, optimal POVMs -- which lead to the saturation of the bounds, may not be realizable experimentally. In our analysis, we have disregarded such limitations except the fact that the QFIMs obtained are non-singular and the attainability criteria, calculated through Eq.~\eqref{eq:attain} is satisfied in all the cases. Therefore, our results are guaranteed to be achievable, in principle, provided the corresponding POVMs are practically possible.

We have shown  that the quantum switch, and the MZ interferometer with composite particle as a thermometer is a tool to measure, and subsequently estimate two temperatures simultaneously. Further, we have identified the quantum control to be an essential element in enabling two-temperature thermometry. Our results underline the role of quantum control -- as an essential asset -- in such multi-parameter estimation setups, particularly enabling interaction of the thermometer with bath(s), such that it is exposed to two temperatures which otherwise seems unfeasible. We can therefore declare conclusively that it is indeed possible to measure two temperatures using a single thermometer, albeit with the assistance of quantum control.

We have shown that though the novel setups introduced here are generally useful, for the MZ setup in particular the determinant of QFIM  is susceptible to the relative phase between its arms. This renders the setup unfeasible for certain values of the aforesaid phase -- those for which the QFIM is singular. In the MZ based setup, we have shown multiple ways in which two-temperature thermometry can be done successfully -- particularly by using a two-qubit probe along with engineering its interaction with the baths, or by using the joint state of a single qubit probe and the control.

In particular, our results give credence to the utility of quantum processes -- particularly quantum switch -- in tasks other than quantum information and computation, which have been widely explored previously. As a further extension of our work, it would be interesting to investigate how other processes, for instance a deterministic quantum process mentioned in \cite{Araujo2017} -- which is known to violate the causal inequality -- fares in the two/three temperature estimation task. Moreover, the utility of time ordered process and the role of non-Markovianity might be interesting to explore for this task. This study could also be complemented by exploring an estimation problem of the squeezing parameter associated with squeezed thermal baths.

In the future, it might also be useful to also contemplate the optimization of input states, and other entangling operations before the measurement, which could lead to a further enhancement in the variances obtained through different setups considered in this work. This study might be also useful in novel quantum machines based on quantum processes, which are connected to two or more thermal reservoirs. However, such machines, including the broader framework of quantum thermodynamics with processes \cite{Capela_2022}, still remain an open area of active theoretical research.

\acknowledgements
The authors thank Carolyn E. Wood, and Magdalena Zych for useful discussions.
F.C.\ acknowledges support through the Australian Research Council (ARC) DECRA grant DE170100712,  F.C.\ acknowledges support through ARC Centre of Excellence EQuS CE170100009.  Nordita is supported in part by NordForsk. The University of Queensland (UQ) acknowledges the Traditional Owners and their custodianship of the lands on which UQ operates.%The authors acknowledge the traditional owners of the land on which the University of Queensland is situated, the Turrbal and Jagera people.

\bibliography{main.bib}

\end{document}